\def\year{2022}\relax
\documentclass[letterpaper]{article} 
\usepackage{aaai22}  
\usepackage{times}  
\usepackage{helvet}  
\usepackage{courier}  
\usepackage[hyphens]{url}  
\usepackage{graphicx} 
\usepackage{natbib}  
\usepackage{caption} 
\usepackage{amssymb}
\usepackage{multirow}
\usepackage{subfigure}
\usepackage{amsmath}
\usepackage{makecell}
\newcommand{\RNum}[1]{\uppercase\expandafter{\romannumeral #1\relax}}
\DeclareCaptionStyle{ruled}{labelfont=normalfont,labelsep=colon,strut=off} 
\usepackage[linesnumbered,ruled,boxed]{algorithm2e} 
\usepackage{algpseudocode}
\usepackage{newfloat}
\usepackage{listings}
\title{MoCA: Incorporating Multi-stage Domain Pretraining and Multimodal \\Cross Attention for Textbook Question Answering}
\author{
    Fangzhi Xu\textsuperscript{\rm 1,2},
    Qika Lin\textsuperscript{\rm 1,2},
    Jun Liu\textsuperscript{\rm 1,3,*},
    Lingling Zhang\textsuperscript{\rm 1,3},
    Tianzhe Zhao\textsuperscript{\rm 1,2},
    Qi Chai\textsuperscript{\rm 1,3},
    Yudai Pan\textsuperscript{\rm 1,3}
}
\affiliations{
    \textsuperscript{\rm 1}Department of Computer Science and Technology, Xi'an Jiaotong University, China\\
    \textsuperscript{\rm 2}National Engineering Lab for Big Data Analytics, Xi’an Jiaotong University, China\\
	\textsuperscript{\rm 3}Shaanxi Province Key Laboratory of Satellite and Terrestrial Network Tech. R\&D, Xi’an Jiaotong University, China
}

\usepackage{bibentry}

\begin{document}

\maketitle

\begin{abstract}
Textbook Question Answering (TQA) is a complex multimodal task to infer answers given large context descriptions and abundant diagrams. Compared with Visual Question Answering(VQA), TQA contains a large number of uncommon terminologies and various diagram inputs. It brings new challenges to the representation capability of language model for domain-specific spans. And it also pushes the multimodal fusion to a more complex level. To tackle the above issues, we propose a novel model named MoCA, which incorporates multi-stage domain pretraining and multimodal cross attention for the TQA task. Firstly, we introduce a multi-stage domain pretraining module to conduct unsupervised post-pretraining with the span mask strategy and supervised pre-finetune. Especially for domain post-pretraining, we propose a heuristic generation algorithm to employ the terminology corpus. Secondly, to fully consider the rich inputs of context and diagrams, we propose cross-guided multimodal attention to update the features of text, question diagram and instructional diagram based on a progressive strategy. Further, a dual gating mechanism is adopted to improve the model ensemble. The experimental results show the superiority of our model, which outperforms the state-of-the-art methods by 2.21\% and 2.43\% for validation and test split respectively.
\end{abstract}

\section{1\quad Introduction}
\noindent Recent years have witnessed the promising development of Visual Question Answering (VQA) task\cite{vqa}, which is required to infer the answers based on an image and its relevant question text. Promoted by the continued researches on multimodal inference, the previously proposed Textbook Question Answering (TQA) task \cite{tqa} leads a new trend. Similarly, the TQA task also requires the model to give the answers based on complex multimodal inputs. Figure 1 illustrates an example of TQA. Two questions are listed on the right part. Different from VQA in general domain, TQA dataset relates to the domain of textbook, containing a large number of span-level terminologies. The evidence spans like $continental$ $slope$ in $Q2$ are crucial to TQA inference, but they are uncommon in the general domain. Meanwhile, the inference of questions relies on the joint consideration of abundant context and diagrams. Take $Q1$ as an instance, the model is required to focus on the span $benthic$ $zone$ in the question and find the most related instructional diagram $ID\raisebox{0mm}{-}2$ in the context. After extracting the corresponding information of $QD\raisebox{0mm}{-}1$ and $ID\raisebox{0mm}{-}2$, the answer can be worked out. From these perspectives, TQA task raises new challenges to multimodal inference.

Firstly, general language model (LM) is insufficient for the specific domain knowledge. Pretraining-based Transformer structures \cite{attention}, such as BERT \cite{bert} and GPT \cite{gpt}, show excellent performance on the general domain. However, TQA context ranges from astrophysics to life science, containing a large number of terminologies in the field of textbook (e.g., $benthic$ $zone$ and $continental$ $slope$ in Figure 1). There exists an obvious gap between specific domain and general domain. Some previous works \cite{dont,selectivemask} have been aware of this situation and adapted LM to some domains (e.g., News, Economics, and Reviews). However, they focus on the token-level information for inference, which fails to concern about span-level terminologies and evidences. To the best of our knowledge, so far there is no attempt to enhance LM for TQA by adding external knowledge or attending to span-level information.

\begin{figure}[t]
	\large
	\centering
	\includegraphics[scale=0.47]{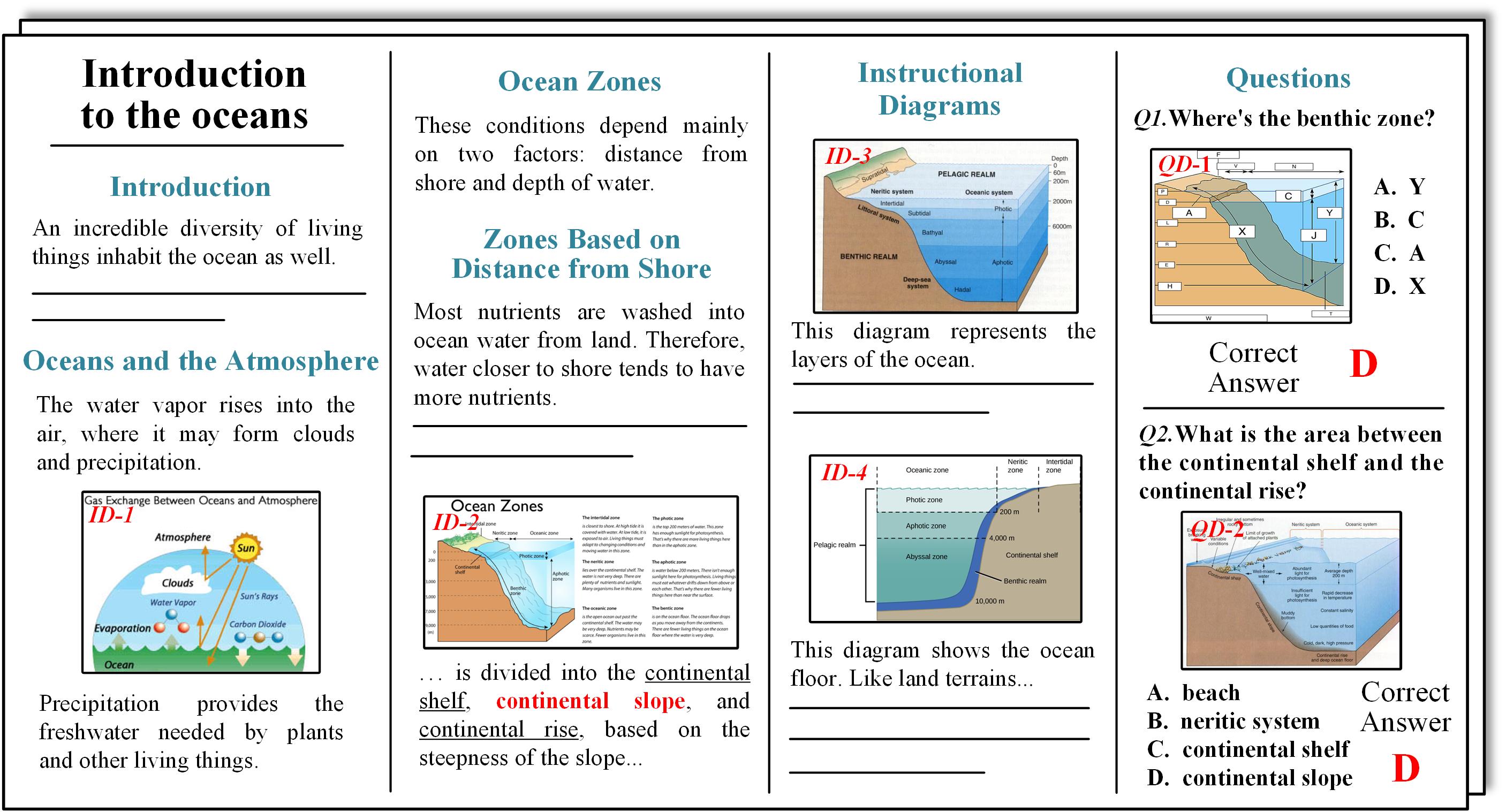}
	\caption{An example of TQA. $ID$ is short for instructional diagram while $QD$ is short for question diagram.}
	\label{fig_dataset}
\end{figure}

Secondly, it is difficult to fully utilize abundant multimodal inputs. Different from VQA with only single natural image, TQA contains various instructional diagrams and one question diagram as visual input. The two types of diagrams are similar in structure and all have complementary information with text (e.g., $QD\raisebox{0mm}{-}1$ and $ID\raisebox{0mm}{-}2$ are closely connected under the guidance of text $benthic$ $zone$ in Figure 1). Although most popular VQA models \cite{butd,mcan,ban} are capable of fusing multimodal features, they lack the ability to update fine-grained features interactively between two diagrams. Several previous works in TQA \cite{isaaq,xtqa,igmn} have noticed the multiple types of diagrams, but they simply process two types in the same way, ignoring their independent effects.

In light of the above challenges, we propose a novel model MoCA, which incorporates multi-stage domain pretraining and cross-guided multimodal attention for the TQA task. We introduce multi-stage pretraining to conduct post-pretraining with the span mask strategy and pre-finetune sequentially between general pretraining-finetune paradigm. Especially for the employment of the terminology corpus, we propose a heuristic generation algorithm. Meanwhile, patch-level diagram representations are obtained through the Vision Transformer \cite{vit}. Then, we construct token-patch pair interaction and obtain attended features based on the multi-head guided attention. Inspired by the attention flow of human inference \cite{DBLP:conf/eccv/ChenJYZ20}, the features of text, question diagram and instructional diagram are updated in a progressive and interactive way. After the final fusion of all the features, the answer is predicted with a dual gating mechanism. The main contributions are shown as follows:

\begin{itemize}
\item A unified model MoCA is proposed to address both the representation for terminologies and the feature fusion of abundant multimodal inputs. We are the first to simultaneously focus on the two challenges in TQA.
\item We introduce a heuristic generation algorithm for terminology corpus. Based on the external knowledge, we are the first to design a span mask strategy in the pretraining stage for the TQA task. 
\item To address the multimodal fusion challenges, a cross-guided attention mechanism is proposed to update the features of rich inputs. In a progressive manner, the interactive updates of the features are obtained.
\item Extensive experiments show that our model significantly improves the state-of-the-art (SOTA) results in the TQA task. Furthermore, ablation and comparison experiments prove the effectiveness of each module in our model.
\end{itemize}

\section{2\quad Related Work}
\textbf{Visual Question Answering.} VQA has aroused wide concerns \cite{DBLP:journals/pami/CaoLLL21, DBLP:conf/sigir/JainKKJRC21, DBLP:conf/acl/Khademi20, DBLP:journals/pr/YuZWZHT20}, as it is regarded as a typical multimodal task related to natural language processing and computer vision. Given an image as well as question text, the model is required to give the answer. Some early models attempted to jointly consider the multimodal inputs. \citet{vqa} encoded the image and text respectively and map them into a common space. \citet{mmresidual} proposed a residual structure to learn joint embeddings. These methods are limited to global and coarse information. 

Therefore, more following works applied the attention mechanism to conduct fine-grained reasoning. Typically, \citet{ban} proposed a bilinear attention network to reduce computational cost. \citet{mutan} proposed a framework to efficiently parametrize bilinear multimodal interactions. \citet{mcan} included the self-attention and the question-guided attention within deep modular co-attention networks. \citet{butd} introduced the bottom-up and top-down attention to attend to object-level and salient information. However, these models are placed in an ideal scenario, which includes unitary input of an image and a short question. They lack the potential to process large context and abundant diagrams.

\noindent\textbf{Textbook Question Answering.} Much attention has been paid to the TQA task since it was proposed. For example, \citet{igmn} aimed to find contradictions between answers and context, and further employ memory network for inference. \citet{fgcn} built a multimodal context graph and introduced open-set learning based on self-supervised method. \citet{rafr} proposed fine-grained relation extraction to reason over the nodes of the constructed graphs. The above three methods focused more on inference process, but ignored the huge potential of multimodal encoding, causing relatively weak representation ability. Under this circumstance, \citet{isaaq} utilized the pretrained transformers for text encoding and bottom-up and top-down attention for multimodal fusion, significantly improving the performance. However, it neglected the span-level knowledge during pretraining and simply treated the question and instructional diagrams in the same way. \citet{xtqa} put the explainability at the first place and attempted to extract useful spans, but its text representation method is relatively weak for the textbook domain. 

Considering the drawbacks of the above models, we apply external knowledge to enhance the span-level representation. Different from them, we treat two types of diagrams respectively to conduct fine-grained feature updates.

\section{3\quad Methods}
In this section, we will introduce our proposed model MoCA. The architecture of MoCA is shown in Figure 2. The left part is an input example of TQA. Then text and diagrams are encoded respectively, through \textbf{M}ulti-stage \textbf{P}retrain (MP) module and \textbf{P}atch-level \textbf{D}iagram \textbf{R}epresentation module (PDR). In PDR module, we employ Vision Transformer to obtain the patch-level representation of both types of the diagrams. Further, multimodal features are progressively updated based on \textbf{C}ross-\textbf{G}uided \textbf{M}ultimodal \textbf{A}ttention (CGMA). Finally, the answers are worked out by \textbf{G}ating \textbf{M}odel \textbf{E}nsemble (GME). The details of three main modules MP, CGMA and GME will be covered as follows. 

\begin{figure*}[t]
	\large
	\centering
	\includegraphics[scale=0.45]{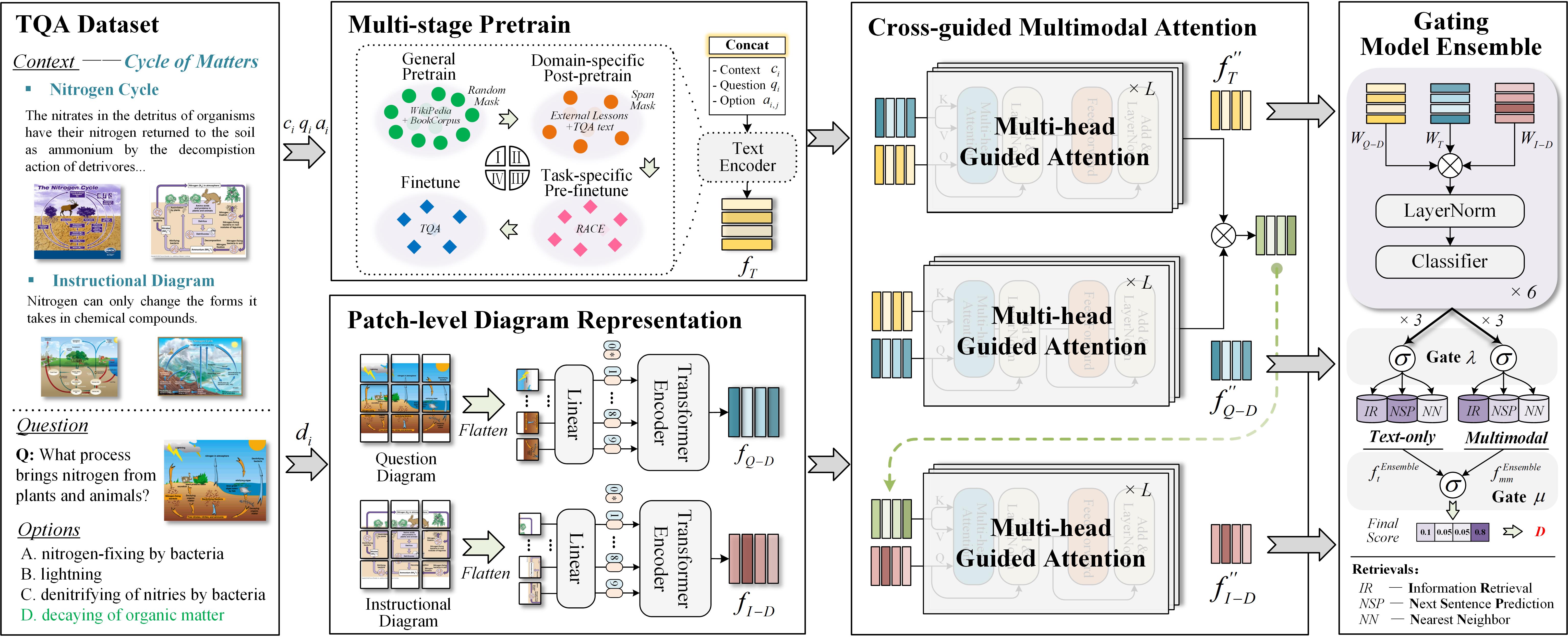}
	\caption{The architecture of MoCA for the TQA task.}
	\label{fig_model}
\end{figure*}

\subsection{3.1\quad Task Formulation}
Given the TQA dataset $\mathcal{D}$ with $M$ questions, the inference of $i^{th}$ question ($i \in [0,M-1]$) can be defined as follows:

\begin{equation}
	\hat{a} =\underset{a_{i,j} \in A_{i}}{\arg \max }  p\left(a_{i, j} \mid c_{i}, q_{i}, A_{i}, d_{i} ; \theta\right),
\end{equation}
where $c_{i}$, $q_{i}$, $A_{i}$ represent the text-only context, question sentence and candidate set respectively. $d_{i}$ includes question and instructional diagrams. The option number in $A_{i}$ is $n$, $j \in [0,n-1]$ and $a_{i,j}\in A_{i}$ represents $j^{th}$ option. $\hat{a}$ is the predicted option. $\theta$ is the trainable parameter.

\subsection{3.2\quad MP Module}
Although the two-stage general pretraining and finetune models has achieved great success in many tasks, there exists an obvious gap between general domain and textbook domain. We attribute the problem to the lack of data about both specific domain and task. Under this circumstance, we propose MP module for the enhancement of text representation.

MP includes four stages in total, which is constructed with two unsupervised pretraining stages and two supervised finetune stages. Stage \RNum{1} and Stage \RNum{4} inherit the traditional general paradigm, with pretraining on large general corpus and finetune on downstream task (TQA) respectively. We use RoBERTa \cite{roberta} as the base model for Stage \RNum{1}, which utilizes the dynamic random mask strategy to focus on token-level performance.

For Stage \RNum{2}, to adapt the language model to the textbook domain, we introduce a coarse-to-fine strategy to heuristically generate external domain corpus. In the beginning, we crawl the textbook-related websites, forming the large coarse-level domain corpus. To evaluate the similarity of the external knowledge with TQA, we employ vocabulary overlap of the Top 1k most frequent words, which excludes nearly 900 stopwords. We define the overlap operator as $\mathcal{O}(\mathbf{c}_{1}|\mathbf{c}_{2})$. It stands for the overlap vocabulary list of input corpus or text $\mathbf{c}_{2}$ with $\mathbf{c}_{1}$, while $L(\mathcal{O}(\mathbf{c}_{1}|\mathbf{c}_{2}))$ returns the number of overlap words. Based on the operator and coarse-level corpus, we design a heuristic method to generate fine-level domain corpus containing rich terminologies. For each line of the corpus, we retain the ones which contribute more to the vocabulary overlaps during every iteration (shown in Algorithm 1, more details are attached to Appendix). After obtaining fine-level corpus, we shuffle it with TQA text to prepare for the post-pretraining stage.

Considering that quite a lot of knowledge and question information consist of multiple words, we adopt span mask strategy to optimize domain-specific pretraining process. That is to say, given a sequence $S=\{t_{1},t_{2},...,t_{n}\}$ consisting of $n$ tokens, each time we randomly mask one to ten tokens based on geometric distribution. Until the mask percentage reaches 15\%, the process breaks automatically. Same as BERT, for 15\% tokens selected in sequence $S$, 80\% of them are replaced with $<$$mask$$>$ flag, 10\% of the tokens are substituted by random token and the others remain unchanged.

\begin{algorithm}[t]
	\KwIn{Coarse-level Corpus $\mathbf{C}_{coarse}$, TQA Text $\mathbf{T}$, threshold $\delta$}
	\KwOut{Fine-level Corpus $\mathbf{C}_{fine}$}
	Calculate vocaburary overlap $\mathcal{O}(\mathbf{T}|\mathbf{C}_{coarse})$ of coarse corpus with TQA text.\\
	\Repeat{$\mathbf{C}_{fine}$ is reduced to a certain specification}
	{
		Take $\mathbf{C}_{fine}$ as $\mathbf{C}_{coarse}$ if not the first iteration\\
		\For{Line $l_{i} \in \mathbf{C}_{coarse}$}
		{
			$\mathbf{W}\leftarrow \mathcal{O}(\mathbf{T}|\mathbf{C}_{coarse})$\\
			$Score(l_{i})\leftarrow \frac{L(\mathcal{O}(\mathbf{W}|l_{i}))}{L(\mathbf{W})}-\frac{L({l_{i}})}{L(\mathbf{C}_{coarse})}$\\
			\If{$Score(l_{i})>\delta$}
			{
				Include $l_{i}$ into fine-level corpus $\mathbf{C}_{fine}$
			}
		}
	}
	\textbf{return} Fine-level Corpus $\mathbf{C}_{fine}$\\
	\caption{Heuristic Terminology Corpus Generation}
\end{algorithm}

For Stage \RNum{3}, the model is trained to adapt to the specific task with the help of an external RACE dataset. Similar to TQA, RACE dataset is a multiple-choice task, which is collected from examinations. Through the process of pre-finetune, the performance of our model is enhanced.

The pretrained MP encoder is used to model the text-only part, including context, question text, and options. For each question, we concatenate the context $c_{i}$, question $q_{i}$ and option $a_{i,j}$ as the input sequence:

\begin{equation}
{\rm Input}(c_{i},q_{i},a_{i,j}) = [C L S] c_{i} [S E P] q_{i} [S E P] a_{i, j} [S E P].
\end{equation}

Feed it into our MP encoder and output the feature of the text $f_{T} \in \mathbb{R}^{N \times d}$, where $N$ represents the max length of the input sequence and $d$ represents the hidden size.

\subsection{3.3\quad CGMA Module}
Through PDR module, the original features of question diagram $f_{Q\raisebox{0mm}{-}D} \in \mathbb{R}^{P \times d}$ and instructional diagram $f_{I\raisebox{0mm}{-}D} \in \mathbb{R}^{P \times d}$ are obtained, where $P$ is patch number and $d$ is the dimension of hidden state. To align text and visual parts to the same space, we add a linear projection layer $\rm Linear(P,N)$. Thus, three features with the same dimension can function as the input of CGMA module. 

\begin{figure}[t]
	\large
	\centering
	\includegraphics[scale=0.43]{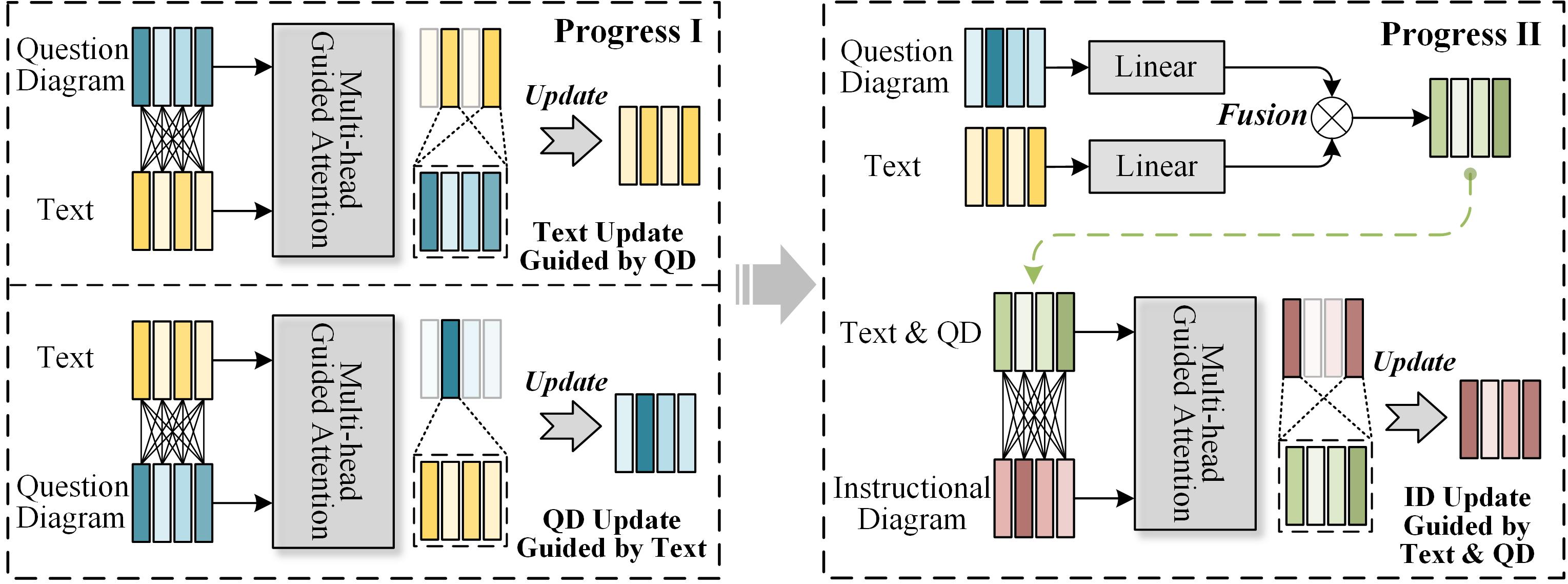}
	\caption{Progressive multimodal feature update in CGMA.}
	\label{fig_model}
\end{figure}

Inspired by the human inference pattern, we propose the progressive attention flow for the TQA task, illustrated in Figure 3. Totally two progresses are included in CGMA. In Progress \RNum{1}, question diagram and text are considered as two inputs, while instructional diagram is added in Progress \RNum{2}. Three types of features are updated with three similar multi-head guided attention. Within each attention, we first build a fine-grained connection between each token-patch pair and focus on the important parts based on attention weights. Take question diagram feature update as an example. The text feature $f_{T}$ is utilized to instruct the feature update of question diagram $f_{Q\raisebox{0mm}{-}D}$. We first map the two features into three matrices $Q_{Q\raisebox{0mm}{-}D}$, $K_{T}$ and $V_{T}$.

\begin{equation}
\begin{aligned}
Q_{Q\raisebox{0mm}{-}D} &= f_{Q\raisebox{0mm}{-}D} \cdot{\rm  W^{Q}},\\
K_{T} &= f_{T} \cdot {\rm W^{K}},\\
V_{T} &= f_{T} \cdot {\rm W^{V}},
\end{aligned}
\end{equation}
where ${\rm W^{Q}, W^{K}, W^{V}} \in \mathbb{R}^{d \times d}$ are projection matrices, and the obtained matrices $Q_{Q\raisebox{0mm}{-}D}, K_{T}, V_{T} \in \mathbb{R}^{N \times d}$. We compute the attention based on the query, key and value matrices.

\begin{equation}
Att(Q_{Q\raisebox{0mm}{-}D}, K_{T}, V_{T}) = {\rm softmax}(\frac{Q_{Q\raisebox{0mm}{-}D}K_{T}^{T}}{\sqrt{d}}) \cdot V_{T}.
\end{equation}

Multi-head guided attention is applied to improve the robustness and capacity of the attention module. The feature can be represented as follows, with the head number $H$.

\begin{equation}
Att_{MH}(Q_{Q\raisebox{0mm}{-}D}, K_{T}, V_{T}) = [Head_{1}, ..., Head_{H}] \cdot {\rm W^{H}},
\end{equation}
where ${\rm W^{H}}\in \mathbb{R}^{(H*d) \times d}$ is the projection matric, $Head_{i}=Att_{i}(Q_{Q\raisebox{0mm}{-}D}, K_{T}, V_{T})$, the input query, key and value matrix are obtained by linear projection of ${\rm W_{i}^{Q}, W_{i}^{K}, W_{i}^{V}} \in \mathbb{R}^{d \times d}$ respectively.

Then, we add the original question diagram features to the computed attention, followed with layer normalization.

\begin{equation}
f_{Q\raisebox{0mm}{-}D}^{\prime} = {\rm LayerNorm}(f_{Q\raisebox{0mm}{-}D}+Att_{MH}(Q_{Q\raisebox{0mm}{-}D}, K_{T}, V_{T})).
\end{equation}

Through the followed feedforward and layer normalization, the attended output feature of one multi-head guided attention layer is obtained. To further enhance the performance of the module, we concatenate $L$ paralleled layers and obtain the multi-layer updated feature $f^{\prime\prime}_{Q\raisebox{0mm}{-}D}$.

\begin{equation}
f^{\prime\prime}_{Q\raisebox{0mm}{-}D} = [Layer_{1}, ..., Layer_{L}] \cdot {\rm W^{L}},
\end{equation}
where ${\rm W^{L}} \in \mathbb{R}^{(L*d) \times d}$ is the projection matric and $Layer_{j}$ represents the output feature $f_{Q\raisebox{0mm}{-}D}^{\prime}\in \mathbb{R}^{N \times d}$ of $i\raisebox{0mm}{-}th$ layer. 

Thus, the feature of question diagram is updated. The text feature can be updated with the similar process in Progress \RNum{1}. Considering that instructional diagram is similar to question diagram in structure and contains meaningful information related to text, we first integrate the multimodal features of text and question diagram and make it a guidance for the instructional diagram feature update in Progress \RNum{2}.

\subsection{3.4\quad GME Module}
To reduce the computational cost, we first shorten the large context into a fixed number of sentences with the help of information retrieval method. In consideration of single retrieval has limitations and may bring unnecessary noise, we perform background retrieval three times in different ways. Following ISAAQ \cite{isaaq}, we name them as $IR$, $NSP$ and $NN$. Since three retrieval methods are independent and contribute differently to the inference, we design a weighted ensemble method to fully exploit the advantages of each retrieval. It means that the final feature $f_{o}^{Ensemble}$ for the prediction consists of three parts with different weights.

\begin{equation}
f_{o}^{Ensemble} = \lambda_{1}f_{o}^{IR} + \lambda_{2}f_{o}^{NSP} + (1-\lambda_{1}-\lambda_{2})f_{o}^{NN},
\end{equation}
where $f_{o}^{Ensemble}, f_{o}^{IR}, f_{o}^{NSP}, f_{o}^{IR} \in \mathbb{R}^{N \times d}$, and weight parameters $\lambda_{1}, \lambda_{2}, 1-\lambda_{1}-\lambda_{2}\in[0,1)$.

Further, for diagram multiple-choice questions, diagrams are considered as an important part of the inference. However, in some cases, the help of diagram features can be negative, or the text-only feature can work out the correct answer. In order to reduce noise brought by diagram features, we integrate the text-only feature for inference in model ensemble. We set a gate parameter $\mu$ to model the importance of text-only features and multimodal features.

\begin{equation}
f_{o}^{GME} = (1-\mu) f_{t}^{Ensemble} + \mu f_{mm}^{Ensemble},
\end{equation}
where $f_{t}^{Ensemble}$ and $f_{mm}^{Ensemble}$ represents the weighted text-only feature and multimodal feature respectively. For each option, we obtain different final features. Through the training of the classifier, the answers can be obtained. 

\section{4\quad Experiments}
In this section, extensive experiments are conducted to compare our model with SOTA methods in TQA. Comparison study and parameter analysis are followed to improve the comprehensiveness of the proposed model.

\subsection{4.1\quad Dataset and baselines}
We conduct the experiments on TQA dataset. Covered in the textbook domain, TQA includes 1,076 subjects, namely Life Science, Earth Physics, etc. It consists of 26,260 questions in the form of True/False (T/F), Text Multiple Choice (T-MC) and Diagram Multiple Choice (D-MC). The number of candidate answers and dataset split are shown in Table 1. Column $2$ to $5$ represent the number of questions while the last column represents the number of candidate answers.

\begin{table}[t]
	\centering
	\begin{tabular}{ccccc}
		\hline
		\textbf{Type} &\textbf{Train Set} &\textbf{Val Set} &\textbf{Test Set} &\textbf{Candidate}\\
		\hline
		T/F & 3,490 &998 &912 & 2\\
		T-MC & 5,163 &1,530 &1,600 & 4-7\\
		D-MC & 6,501 &2,781 &3,285 & 4\\
		\hline
	\end{tabular}
	\caption{Dataset split and candidate answer number}
	\label{tab:TQA}
\end{table}

To prove the superiority of our model, we employ the following baselines, including the SOTA model. 


\begin{itemize}
\item \textbf{Random}: Results based on random prediction.

\item \textbf{MemN} \cite{tqa}: It employs the concept of memory network and diagram parse graph to construct the context graph for the TQA task.

\item \textbf{IGMN} \cite{igmn}: It grasps the contradiction between candidate answers and context for reasoning.

\item \textbf{FCC} \cite{fcc}: It jointly considers diagrams and image captions.

\item \textbf{f-GCN1} \cite{fgcn}: A new f-GCN module based on graph convolution network is proposed to address multimodal fusion challenges.

\item \textbf{XTQA} \cite{xtqa}: It designs a coarse-to-fine algorithm to generate span-level evidences. 

\item \textbf{RAFR} \cite{rafr}: A fine-grained reasoning network is proposed to reason over the nodes of relation-based diagram graphs.

\item \textbf{ISAAQ} \cite{isaaq}: It improves the SOTA baseline by introducing the pretrained LM and bottom-up and top-down attention mechanism.

\end{itemize}

\begin{table}[t]
	\centering
	\begin{tabular}{cccc}
		\hline
		\textbf{Corpus} & \textbf{Domain} & \textbf{Size} & \textbf{Source}\\
		\hline
		TQA text & Textbook & 5MB & From TQA\\
		External & Textbook & 350MB & Crawled\\
		\hline
		\textbf{Dataset} & \textbf{Domain} & \textbf{Size} & \textbf{Type}\\
		\hline
		TQA & Textbook &26,260  & Multimodality\\
		RACE & Exams &97,687  & Text-only\\
		VQA(Part) & General &90,000  & Multimodality\\
		AI2D & Science &8,730  & Diagram-only\\
		\hline
	\end{tabular}
	\caption{The detailed information of corpus and dataset employed in the training process.}
	\label{tab:corpus}
\end{table}

\subsection{4.2\quad Implementation Details}
\begin{table}[t]
	\centering
	\begin{tabular}{m{1.75cm}<{\centering}ccccc}
		\hline
		Model & T/F & T-MC & T-All & D-MC & All\\
		\hline
		Random &50.10&22.88&33.62&24.96&29.08 \\
		MemN &50.50&30.98&38.69&32.83&35.62 \\
		IGMN &57.41&40.00&46.88&36.35&41.36 \\
		FCC &-&36.56&-&35.30&-\\
		fGCN &62.73&49.54&54.75&37.61&45.77\\
		XTQA &58.24&30.33&41.32&32.05&36.46\\
		RAFR &53.63&36.67&43.35&32.85&37.85\\
		ISAAQ &\underline{81.36} &\underline{71.11} &\underline{75.16} &\underline{55.12} &\underline{64.66}\\
		MoCA(Ours) &\textbf{81.56} &\textbf{76.14} &\textbf{78.28} &\textbf{56.49} &\textbf{66.87}\\
		\hline
	\end{tabular}
	\caption{Experimental results on the validation split for TQA. The percentage signs (\%) of accuracy values are omitted. The optimal and suboptimal results are marked in bold and underline respectively (same for the following tables).}
	\label{yab:performance_val}
\end{table}

\begin{table}[t]
	\centering
	\begin{tabular}{m{1.75cm}<{\centering}ccccc}
		\hline
		Model & T/F & T-MC & T-All & D-MC & All\\
		\hline
		XTQA &56.22&33.40&46.73&33.34&36.95\\
		RAFR &52.75&34.38&41.03&30.47&35.04\\
		ISAAQ &\underline{78.83} &\underline{72.06} &\underline{74.52} &\underline{51.81} &\underline{61.65}\\
		MoCA(Ours) &\textbf{81.36} &\textbf{76.31} &\textbf{78.14} &\textbf{53.33} &\textbf{64.08}\\
		\hline
	\end{tabular}
	\caption{Experimental results on test split for TQA.}
	\label{tab:performance1}
\end{table}

All of the experiments are finished with a single GPU of Tesla V100. As for the encoder of text in MP module, we utilize the RoBERTa-large model for Stage \RNum{1} which has 1024-dimensional embeddings. For Stage \RNum{2}, we mix the external textbook corpus with TQA corpus and conduct the post-pretraining with 10 epochs based on the span mask strategy. For Stage \RNum{3}, we pre-finetune the model on RACE dataset with 4 epochs and select the best one based on the performance of test split. As for the background information retrieval, we follow the three methods introduced in ISAAQ \cite{isaaq} namely $IR$, $NSP$ and $NN$. Considering the length of retrieved context, we set the maximum input sequence length to 180, and the inputs shorter than 180 are padded to max. As for diagram representation, each diagram is cut into 14$\times$14 patches with the dimension of 1024. As for CGMA module, we empirically set the multi-head number to 8 and the number of paralleled layers is searched for the best in \{1,2,3,4\}. To enhance the classifier for D-MC questions, we further finetune the model on VQA \cite{vqa} and AI2D \cite{ai2d}. The detailed information of the corpus and dataset is shown in Table 2.

\subsection{4.3\quad Main Results}
MoCA model is evaluated on TQA dataset. The results on validation and test split are shown in Table 3 and Table 4 respectively. Since some previous baselines do not make the results of test split public, we only compare the validation split results for these models.

From the results, MoCA outperforms the SOTA results from all three types of questions. In general, we significantly improve the overall performance of SOTA by 2.21\% and 2.43\% on validation and test split respectively. It is worth mentioning that MoCA also shows better generalization capability. Especially for the text-only questions on the test split, MoCA outperforms the previous SOTA method by 3.62\%. Also, results on previous models illustrate the different distributions between validation and test split for D-MC questions, while MoCA narrows this obvious gap.

\subsection{4.4\quad Ablation and Comparison Experiments}
\begin{table}[t]
	\centering
	\begin{tabular}{p{3cm}ccc}
		\hline
		Model(Valid) & T/F & T-MC & T-All\\
		\hline
		MoCA &81.56 &76.14 &78.28\\
		\quad w/o Stage \RNum{2} &79.86 &75.03 &76.94\\
		\quad w/o Stage \RNum{3} &79.76 &75.88 &77.41 \\
		\quad w/o Stage \RNum{2} \& \RNum{3} &79.26 &75.16 &76.78 \\
		\quad w/o LSOs &81.16 &73.73 &76.66 \\
		\hline
		\hline
		Model(Test) & T/F & T-MC & T-All\\
		\hline
		MoCA &81.36 &76.31 &78.14\\
		\quad w/o Stage \RNum{2} &78.07 &75.19 &76.24\\
		\quad w/o Stage \RNum{3} &78.51 &75.56 &76.63 \\
		\quad w/o Stage \RNum{2} \& \RNum{3} &78.73 &74.00 &75.72 \\
		\quad w/o LSOs &80.92 &72.88 &75.80 \\
		\hline
	\end{tabular}
	\caption{Ablation experiments for MP module}
	\label{tab:ablationMP}
\end{table}

\begin{table}[t]
	\centering
	\begin{tabular}{p{3.2cm}cccc}
		\hline
		Model(Valid) & IR & NSP &NN &T-MC\\
		\hline
		MoCA &73.33 &71.83 &68.17 &76.14\\
		\quad - Random Mask &72.09 &71.11 &68.04 &75.36  \\
		\quad - Whole Word Mask &72.09 &71.63 &68.69 &76.08  \\
		\hline
	\end{tabular}
	\caption{Comparison results of mask strategy in MP module.}
	\label{tab:mask}
\end{table}

\begin{table}[t]
	\centering
	\begin{tabular}{p{2.8cm}cccc}
		\hline
		Model (Valid) & IR & NSP & NN & D-MC\\
		\hline
		MoCA &54.15 &53.83 &51.60 &56.49\\
		\quad w/o ID-Guided &53.72 &53.72 &50.74 &56.17\\
		\quad w/o QD-Guided &53.94 &54.01 &51.06 &56.27\\
		\quad w/o Text-Guided &53.65 & 52.68& 51.13 & 55.63\\
		\quad w/o All Attention &53.15 &53.36 &50.99 &55.02\\
		\quad w/o ID &53.94 &53.72 &51.31 &56.09\\
		\hline
		\hline
		Model (Test) & IR & NSP & NN & D-MC\\
		\hline
		MoCA &52.12 &51.87 &51.26 &53.33\\
		\quad w/o ID-Guided &51.69 &51.93 &51.29 &52.66\\
		\quad w/o QD-Guided &51.57 &51.78 &50.75 &53.12\\
		\quad w/o Text-Guided &51.63 &51.32 & 50.17&53.18 \\
		\quad w/o All Attention &51.08 &51.32 &51.29 &53.15\\
		\quad w/o ID &51.78 &51.11 &51.14 &52.94\\
		\hline
	\end{tabular}
	\caption{Ablation experiments for CGMA module.}
	\label{tab:ablationCGMA}
\end{table}

\begin{table}[h]
	\centering
	\begin{tabular}{p{4.8cm}cc}
		\hline
		Model & Valid & Test\\
		\hline
		MoCA &56.49 &53.33\\
		\quad w/o $\mu$ effect (Three models) &55.45 &52.85\\
		\quad w/o $\mu$ and $\lambda$ (Single model) &54.15 &52.12\\
		\quad w/o $\mu$ and $\lambda$ (Six equal models) &55.88 &53.27\\
		\hline
	\end{tabular}
	\caption{Ablation experiments for GME module.}
	\label{tab:ablationGME}
\end{table}

As MoCA includes three main modules, we conduct the ablation experiments for each module to explore the effectiveness. Meanwhile, comparison experiments on the mask strategy in MP module are also presented.

\noindent\textbf{MP module.} We mainly experiment and analyze the effectiveness of multiple stages and the mask strategy in MP module. Firstly, we remove one or both of Stage \RNum{2} and Stage \RNum{3}. Without Stage \RNum{2} \& \RNum{3}, the text-only accuracy drops by 1.5\% and 2.42\% on validation and test split respectively. Within the removed two stages, Stage \RNum{2} contributes most to the model performance. It proves that the unsupervised training on external knowledge makes a difference.

Through exploratory data analysis, we also discover the characteristics of the options, which is the Latent Semantic Option(LSO). For example, the option `All of the above' has the further meaning that the answer includes the information of all options, which is hard to be reflected by LM. We divide them into two types: Postive and Negative. For the Positive type, like `All of these', `Both a and b', it means the final answer includes more than one option. We concatenate the options and replace the LSO with the spliced text. For the Negative one, like `None of these', `Neither a nor b', we set LSO to the empty string for simplicity. We also test the necessity of LSOs. The evaluation results are shown in Table 5. The consideration of LSOs brings 1.62\% and 2.34\% accuracy gain on validation and test split. From the results, the mapping of LSOs has equal contributions with MP. The former one improves the performance by manually designed rules while the latter one relies on the external knowledge.

Secondly, we select two popular mask strategies as comparison objects, that is random mask and whole word mask. Results on Table 6 show the superiority of the span mask strategy, especially on the retrieval of $IR$ with 1.24\% accuracy gain. Since pretraining on general domain (Stage \RNum{1}) utilizes the random mask strategy rather than span-level strategy, it still remains huge potentials for the improvement.

\noindent\textbf{CGMA module.} We remove one or all of the cross-guided attention. Since previous SOTA models seldom take ID into consideration or just simply confuse it with QD, we also remove the input of ID to show its effectiveness. The experimental results in Table 7 show that all the attention employed brings 1.47\% gain on the validation split. The consideration of ID improves the performance by 0.40\% and 0.49\% on the validation and test split respectively. Further, the update of text feature contributes most on the validation split and least on the test split. It shows the importance of text for D-MC in each split.

\noindent\textbf{GME module.} Our ensemble method has two gate parameters $\lambda$ ($\lambda_{1},\lambda_{2}$) and $\mu$, which has different type of roles. Therefore, we further compare three conditions. Firstly, utilize the multimodal features only for D-MC inference, which eliminates the effect of gate parameter $\mu$ ($\mu$=0.5). Secondly, simultaneously eliminate all the gate parameters, which means select the best single one as the final model ($\lambda,\mu$ not exist). Thirdly, keep both of the gate parameters while set all six models to equal contributions ($\lambda_{1}$=$\lambda_{2}$=1/3,$\mu$=0.5). Results are shown in Table 8. Generally, two gate parameters $\lambda$ and $\mu$ contribute 2.34\% and 1.21\% to the accuracy on the validation and test split respectively. On one hand, more models for ensemble bring better overall performance. On the other hand, single model of MoCA is still competitive to ensembled SOTA model.

\begin{figure}[t]
	\large
	\centering
	\includegraphics[scale=0.50]{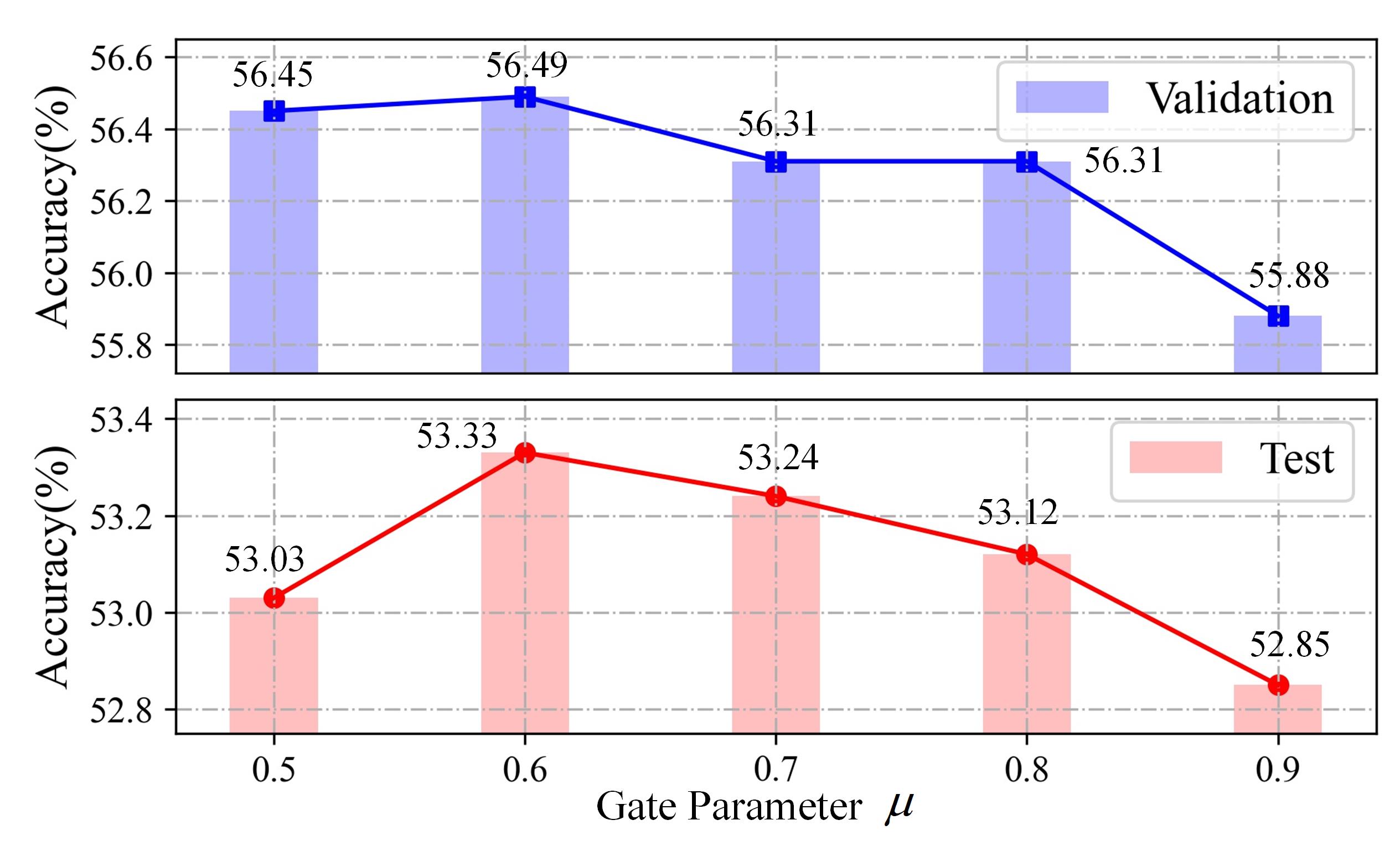}
	\caption{Analysis of gate parameter.}
	\label{fig_gate}
\end{figure}

We select several $\mu$ with an interval of 0.1 for visualization. The results on validation and test split are shown in Figure 4. The trend illustrates that MoCA reaches the best performance when the gate parameter $\mu$ is 0.6. With $\mu$ increasing, accuracy on the validation and test split witnesses an obvious drop. More specifically, the inference of TQA task relies on text-only evidences. It also proves the effectiveness and necessity of GME module.

\subsection{4.5\quad Case Study}

\begin{figure}[t]
	\large
	\centering
	\includegraphics[scale=0.35]{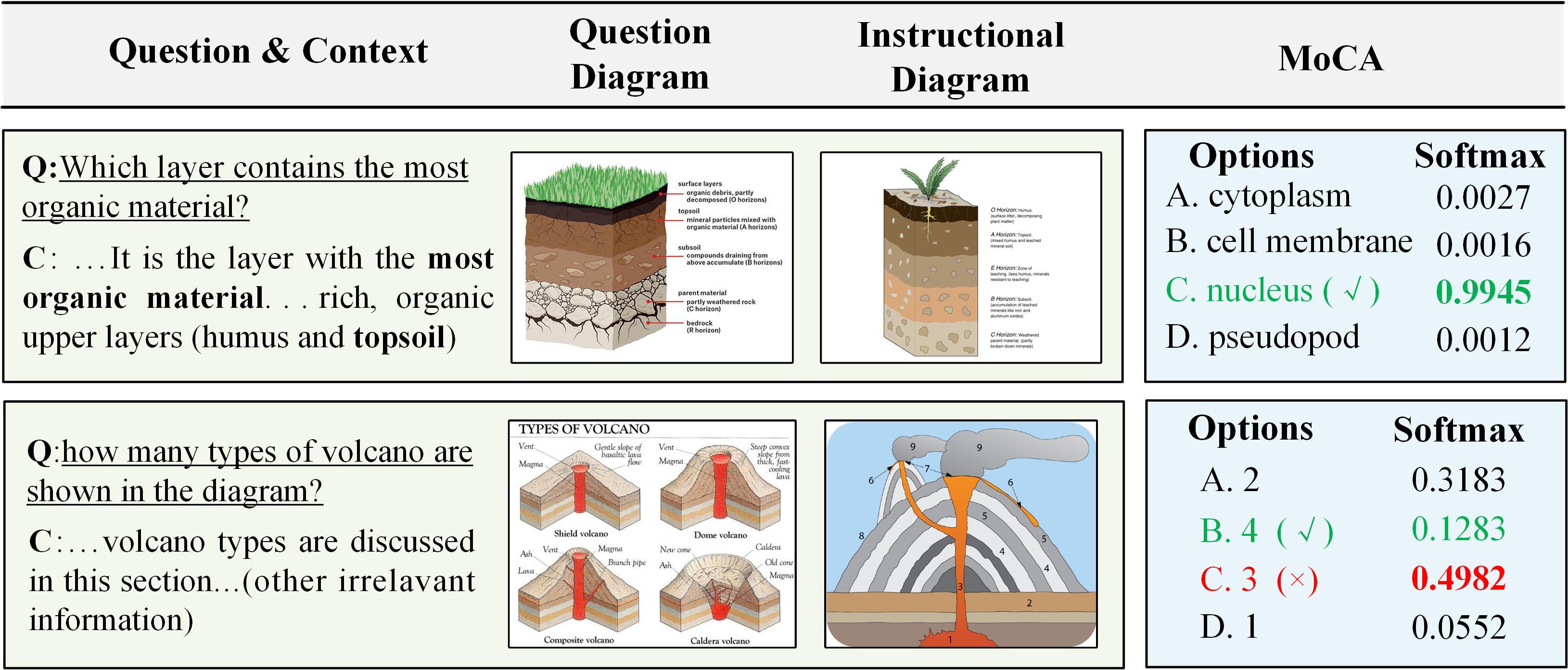}
	\caption{Case study.}
	\label{fig_case}
\end{figure}

\begin{figure}[t]
	\large
	\centering
	\includegraphics[scale=0.225]{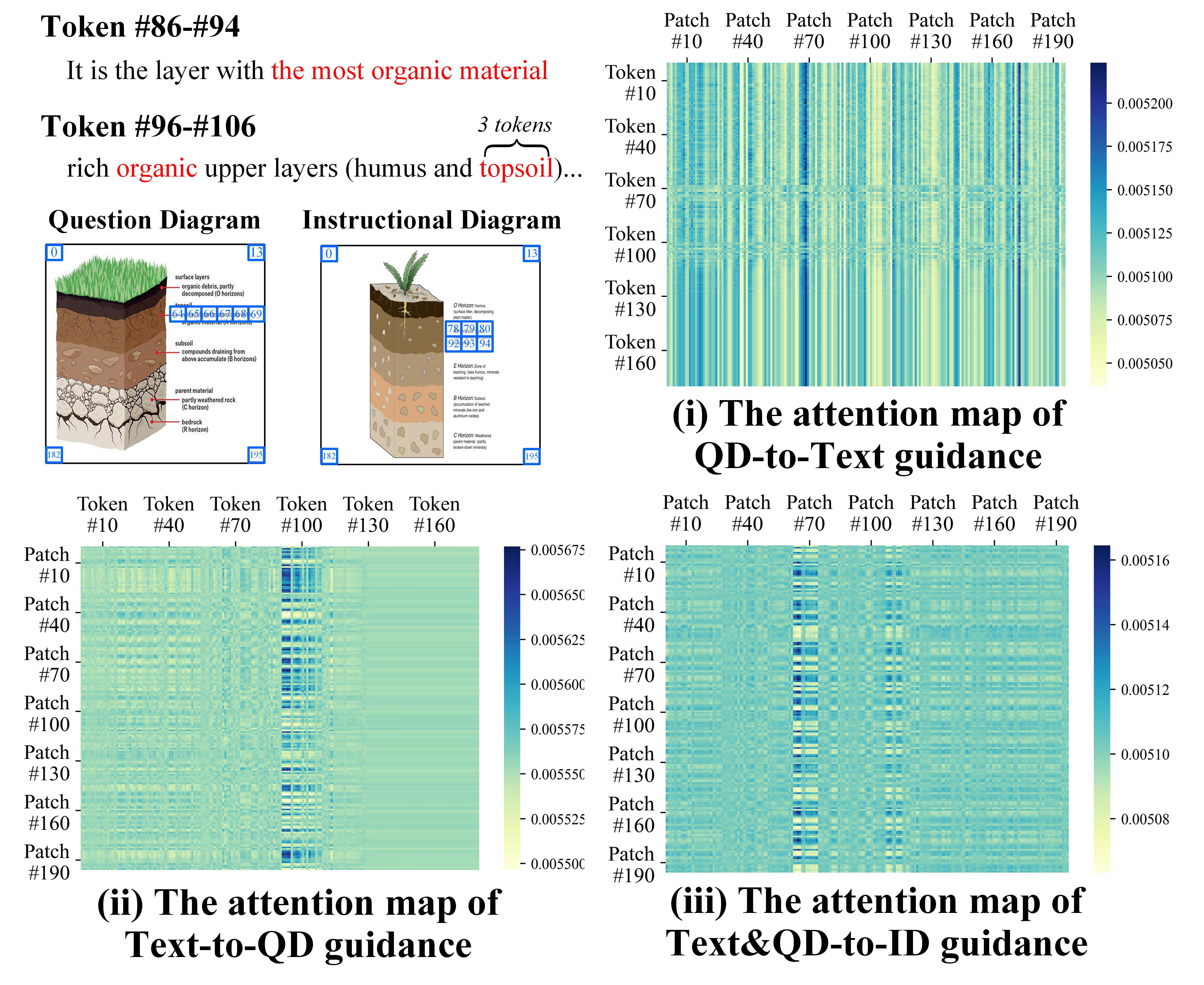}
	\caption{Visualization of attention maps. For better correspondence, we extend the sequence length of diagrams from 180 to 196, with the help of 2-D linear interpolation fitting.}
	\label{fig_QA}
\end{figure}


As text-only questions can be regarded as a special case of diagram questions, we conduct the case study on the diagram questions only. Figure 5 shows a successful case and a failure case. For the successful one, MoCA makes good use of multimodal information for inference. For the failure one, text information is limited and it is required to figure out the number of volcano type directly from QD. It reflects that MoCA is weak in counting and numerical reasoning.

We select the successful case to visualize the effects of three cross-guided attention in MoCA, shown in Figure 6. For better visualization, we mark some important patches with blue boxes in QD and ID, and present the order number above. The left right presents the attention weight of QD-to-text guidance. Obviously from the attention map, Patch \#64 to Patch \#69 contribute most to the attention weight. These patches cover the crucial clues for the correct answer `Topsoil' in QD. The bottom left is the attention weight of text-to-QD guidance. We notice that tokens around \#100 are mainly focused on. These tokens consist of key sentence like `It is the layer with the most organic material' and key span like `rich organic upper layers (humus and topsoil)', which are extremely close to the answer. The bottom right part reflects the attention weight of multimodal guidance to ID. Its X-axis represents the multimodal sequence while Y-axis represents the ID sequence. From the correspondence, important patches of \#78 to \#80 and \#92 to \#94 obtain larger attention weights. The final softmax result after the classifier also proves the positive effect of the cross-guided attention.

\section{5\quad Conclusion}
We incorporate multi-stage domain pretraining and multimodal cross attention for the TQA task. Firstly, on the basis of general pretraining-finetune paradigm, we propose multi-stage domain pretraining module to bridge the gap between general domain and textbook domain. In the stage of domain post-pretraining, we propose a heuristic generation algorithm to employ terminology corpus. Span mask strategy is utilized to optimize the pretraining performance. Secondly, following the human inference pattern, we propose multimodal cross-guided attention to progressively update the features of text, question diagram and instructional diagram. Further, we adopt a dual gating mechanism to improve the ensembled model performance. Extensive experiments prove the superiority of our model and module effectiveness. In the future, we will pay more attention to the improvement of background text retrieval, as well as the employment of fine-grained diagram information.

\bibliography{aaai22.bib}

\end{document}